# Mapping the conformations of biological assemblies

**P Schwander**[1], **R Fung**[1+], **G N Phillips, Jr**[2], **and A Ourmazd**[1*]

[1] Department of Physics, University of Wisconsin Milwaukee, 1900 E. Kenwood Blvd, Milwaukee, WI 53211, USA
[2] Department of Biochemistry and the Center for Eukaryotic Structural Genomics, University of Wisconsin Madison, 433 Babcock Drive, Madison, WI 53706, USA

Email: ourmazd@uwm.edu

**Abstract.** Mapping conformational heterogeneity of macromolecules presents a formidable challenge to X-ray crystallography and cryo-electron microscopy, which often presume its absence. This has severely limited our knowledge of the conformations assumed by biological systems and their role in biological function, even though they are known to be important. We propose a new approach to determining to high resolution the three-dimensional conformations of biological entities such as molecules, macromolecular assemblies, and ultimately cells, with existing and emerging experimental techniques. This approach may also enable one to circumvent current limits due to radiation damage and solution purification.

---

[+] Present address: Department of Physics & Astronomy, University of California Los Angeles, 475 Portola Plaza, m/c 154705, Los Angeles, CA 90095, USA.
[∗] Author to whom correspondence should be addressed.

# 1. Introduction

The biological functions of macromolecules and their assemblies almost always involve conformational changes, which can be quite pronounced or very subtle. High resolution determination of macromolecular conformations is thus an important scientific frontier. In principle, single-particle approaches are ideally suited to determining conformations of biological entities such as molecules, molecular assemblies, chromosomes, and cells. In reality, however, single-particle techniques often rely on "averaging" data obtained from a large ensemble of objects usually assumed to be identical. Conformational information on biological systems thus remains incomplete. Nonetheless, there is increasing appreciation that the dynamic behavior of macromolecules is an inherent part of their functional design. Examples range from the classical descriptions of the R to T states in hemoglobin and related atomic motions in myoglobin [1] to the recent recognition that the virulence of the dengue virus strongly depends on transitions in its protein contacts and conformational rearrangements [2] .

Despite powerful contributions to the study of proteins and some assemblies, X-ray crystallography and NMR have limitations. With notable exceptions, the constraints imposed by crystals have limited the role of X-ray crystallography in elucidating conformational variety. NMR, while able to study conformations in biomolecules of modest size, has not been extensively applied to larger systems. Cryo-Electron Microscopy (cryo-EM) has been extensively used to study nominally identical macromolecular assemblies [3-6]. When conformational variety has been explicitly addressed, the results, won with effort and ingenuity, provide tantalizing evidence of a rich variety of conformations, even in well-studied systems [5-7]. A deep understanding of the nature and role of conformational variety in biological function would revolutionize our knowledge of key processes ranging from basic cell function to pathological states. Unraveling the role of conformations in virulence, for example, is expected to lead to new strategies for fighting infection.

Here, we show that a new generation of algorithms combining techniques from Riemannian geometry, General Relativity, and machine learning is poised to deliver a powerful new approach to biological structure determination in a way which maps conformational heterogeneity, *and* circumvents longstanding limits due to radiation damage and noise. These algorithms can be used with established experimental techniques, such as cryo-EM, and emerging approaches exploiting the extreme brightness of X-ray Free Electron Lasers (XFEL's). By transforming the limits set by the nature of biological entities - weak scattering, radiation damage, conformational variety - to limits set by computational resources, which have improved exponentially for fifty years, these algorithmic approaches promise a decisive advance in biostructure determination.

In the long term, these approaches are expected to constitute a new platform for determining structure and conformations in ways which mitigate the limits set by the nature of biological entities, *and* by vexing experimental issues such as solution purification and crystallization. The work described in this paper is the next vertical step along a continuum beginning with our demonstration that the structure of individual macromolecules can be determined to high resolution from an ensemble of coherent diffraction snapshots of unknown orientation at mean photons counts as low as $\sim 10^{-2}$ per Shannon pixel [8] (figure 1). This is orders of magnitude below the signal levels previously required [9]. The algorithms we have developed have closed a critical gap in proposed techniques for determining the structure of individual biological entities, whereby a succession of *identical* particles is exposed to single, short, and intense pulses from an XFEL source [10,11]. We are now in a position to address the next step: the hitherto unanticipated possibility of determining to high resolution the three-dimensional (3D) structure of the conformations of an object from random snapshots of an ensemble of *non- identical* objects each in a different conformation. This is expected to have a substantial impact on existing cryo-EM techniques [3] and the proposed XFEL-based "scatter-and-destroy" approaches [11], significantly advancing the tomographic study of increasingly complex systems such as chromosomes and perhaps whole cells.

This paper is organized as follows. Section 2 presents a brief overview of recent trends in determining macromolecular structure and conformations. Section 3 outlines a new algorithmic approach to determining structure, with section 4 summarizing our current understanding of its capabilities. Section 5 addresses the application of this approach to determining macromolecular conformations. Section 6 places the various algorithms in context, indicating possible routes to further progress. Section 7 summarizes and concludes the paper.

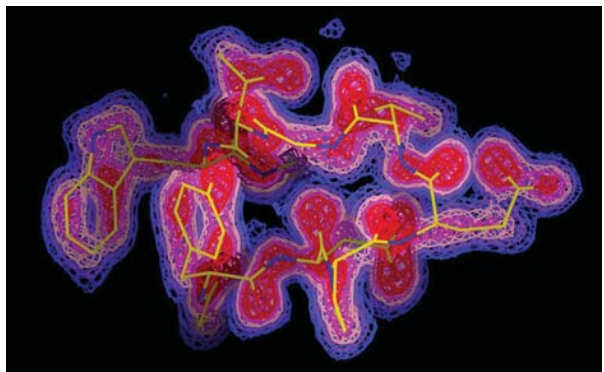

**Figure 1. Structure of small protein chignolin recovered from 72,000 simulated diffraction patterns of unknown orientation at a mean photon count of $4 \times 10^{-2}$ per pixel**. Stick figure is molecular model. C: yellow; N: blue; O: red. 1, 2, and 3σ electron density contours are blue, pink, and red, respectively. (σ: RMS deviation from the mean electron density.)

## 2. Overview of Current Knowledge and Techniques

There is increasing awareness that structural variability is not only common, but can play a key role in macromolecular function. Revealing structural variability in macromolecules and their assemblies often requires spatial resolutions beyond the reach of current X-ray tomographic techniques [12]. Applying sophisticated algorithmic approaches to cryo-EM data, Scheres et al. revealed structural variability in 70S *E. coli* ribosome particles [5] and other well-characterized systems [13]. Combining normal mode analysis with cryo-EM, Brink et al. highlighted the conformational variations of human fatty acid synthase [14]. Yu et al. showed that reversible, pH-driven conformational changes of flaviviruses are central to the mechanism by which they are processed and stabilized in the host cell [2]. A high-resolution cryo-EM study of GroEL revealed significant deviations from existing crystal structures [6]. Even in crystals, a new treatment of data has revealed statistically significant evidence for the presence of an ensemble of conformations [15]. Indeed, sub-nanosecond time-resolved crystallography of conformational changes can be performed [1]. Despite mounting evidence of its importance, the study of structural variability has proved difficult, limiting our ability to relate structure and dynamics to function.

The ultimate quality of experimental structural data and hence achievable spatial resolution are determined by radiation damage and/or noise. Cryo-EM, the single-particle technique of choice, for example, has yielded a host of valuable information, but is severely limited by radiation damage [16]. Emerging XFEL methods have recently used intense short pulses containing up to $10^{12}$ photons to obtain coherent diffraction snapshots of individual particles before the particle is destroyed [17]. By collecting data before significant damage has occurred, these so-called "scatter-and-destroy" approaches [10,18-22] promise to mitigate radiation damage. Several experimental capabilities are needed for this promise to be realized. These include maintaining the native state of biomolecular assemblies injected into the X-ray beam, collecting data with sufficient signal-to-noise ratio before significant radiation damage, and the availability of robust algorithms for reconstructing 3D structure from noisy 2D snapshots of unknown orientation in the presence of background scattering. Recent experimental results obtained at the FLASH soft-X-ray FEL in Hamburg, Germany [22-25] have shown that diffraction snapshots of single biological particles can be obtained in their native state, demonstrating that the necessary experimental techniques are within reach. Simulations indicate the effect of radiation damage to be below the 0.3nm level [26]. The imminent extension of this capability to the hard X-ray regime offers an unprecedented opportunity to determine the 3D structure of macromolecular assemblies to high resolution. The world's first hard-XFEL, the Linear Collider Light Source (LCLS) at the Stanford Linear Accelerator Center (SLAC) produced first photons in April 2009, and commenced lasing 10 days later. There are thus at least two single-particle approaches in principle able to study structural variability at high resolution. They are both affected by radiation damage.

The combination of radiation damage, weak scattering, unknown snapshot orientations, and structural variability constitutes a formidable challenge. Cryo-EM and XFEL approaches can require ~$10^6$ 2D snapshots from identical copies of a biological object to reconstruct its 3D structure. Even in the absence of structural variability, determining the orientations is a key challenge, because the signal-to-noise ratio in each snapshot is so poor [3,9]. The investigation of structural variability, be it due to different conformations or ligand binding states in the ensemble of particles, further complicates matters [3,5]. The study of structural variability by cryo-EM has relied on data from other techniques and/or ad hoc assumptions. Supervised classification, the most commonly used method, sorts the snapshots according to similarity to reference templates, and thus requires prior knowledge of the number and types of structural classes present [27,28]. A recent statistically principled but computationally expensive Expectation-Maximization-based study of structural variability had to resort to trial-and-error to estimate the number of conformations [5]. Nonetheless, this study highlighted the power of algorithmic approaches, which naturally treat structural and orientational variability on an equal footing, and exploit the information content of the entire dataset at each step. It has been proposed to use as-yet unavailable experimental capabilities, such as simultaneous recording of multiple projections with femtosecond accuracy , and/or undemonstrated algorithms to recover the 3D structure of non-identical objects to limited resolution [29,30]. The possibility of determining the 3D structure of conformationally heterogeneous objects to high resolution by XFEL methods is new [8].

The power - and resolution limit - of any reconstruction approach is determined by its ability to extract information from the noisy dataset. Determining the snapshot orientations is perhaps the most critical steps in 3D structure recovery, because it has to be performed at extremely low "raw-signal" levels. Different techniques must therefore be judged by the lowest signal at which they can determine snapshot orientations, in the first instance in the absence of structural variability. Cryo-EM snapshots can be oriented down to a mean electron count of ~10 /$Å^2$, with 30 representing a typical value [6]. The presence of symmetry is often exploited, increasing the effective electron count by the number of symmetry elements, which can be as high as 120 for icosahedral particles viewed in diffraction. Shneerson et al. showed that orienting XFEL coherent diffraction snapshots by the "common-line" approach [31] requires ~1000x more signal than available [9]. The group of this corresponding author published the first demonstration of structure recovery from simulated XFEL snapshots of macromolecules [8]. A second demonstration [32] was recently published using an approach fundamentally similar to that used by us [33]. As in [13], key to success is the realization that the information content of the entire dataset must be used at each step. This is because each snapshot contains information about every other, much as the picture from the back of a person's head provides information about the position of the ears, and thus contributes to reconstructing a full-frontal image. This approach was used to reconstruct the 3D structure of a small macromolecule from simulated XFEL snapshots of unknown orientation [8] (figure 1). Using a simplified model, Elser has argued that the type of approach we have used is capable of operating at even lower signal levels [34]. Approaches exploiting the information content of the entire dataset extract signal from noise with extreme efficiency, pointing the way to 3D structure recovery to unprecedented resolution with established and emerging single-particle techniques.

## 3. New Approach: Manifold Mapping

Of the several approaches for extracting information from the entire dataset, those based on the concept of manifolds are particularly powerful. To appreciate the concept, consider an object able to assume any orientation in 3D space, with each snapshot stemming from an unknown orientation of the object. A snapshot consisting of $p$ pixels can be represented as a $p$-dimensional vector, with each component representing the intensity value at a pixel (figure 2). The fact that the intensities are a function of only three orientational parameters ("Euler angles") means that the $p$-dimensional vector tips all lie on a 3D manifold in the $p$-dimensional space of intensities. This manifold, which represents the information content of the dataset, is traced out by the correlated way in which the $p$ pixel intensities change with particle orientation. Each point on the manifold represents a snapshot at a particular orientation. Determining this manifold allows one to assign an orientation to each snapshot [8]. A number of powerful techniques have been developed to discover low-dimensional manifolds in high-dimensional data [35-38]. Each has its strengths and limitations, with the most common problem being noise sensitivity [39,40]. Some manifold mapping techniques attempt to determine the underlying dimensionality of the manifold from the data, but suffer from the disadvantage that the manifold topology and dimensionality can be strongly affected by noise [39]. We have developed noise-robust versions of Generative Topographic Mapping (GTM) [8,35,41], Isomap [36], and Diffusion Map [38], demonstrating structure recovery with GTM [8] and Diffusion Map and Isomap [42] at $\sim 10^{-2}$ photon/Shannon-pixel. GTM is computationally the most expensive, but has the advantage of allowing one to specify the key variables of the problem (in this case the "Euler angles") as dimensions of a so-called "latent" space, which is then embedded in the "manifest" space of the data.

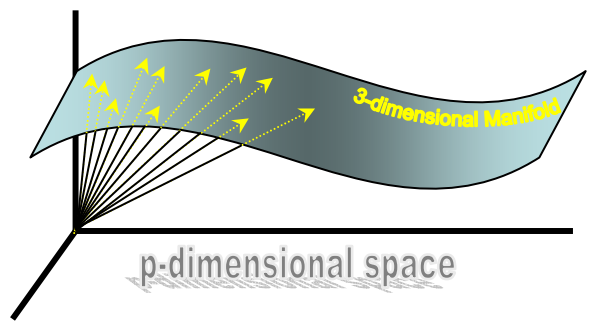

**Figure 2. The manifold expresses the information content of the dataset.** A molecule has only three orientational degrees of freedom. This means that the $p$ pixel intensities in a snapshot change in a correlated fashion with molecular orientation. This correlation is described by a 3D manifold in the $p$-dimensional space of pixel intensities.

The achievable resolution for structure recovery depends on noise, type of algorithm, and available computational resources. Using simulated snapshots from a set of identical objects in unknown orientations, and assuming GTM running on a 100-node cluster of 2.33GHz Intel Core 2 Duo processors, we have shown that it should be possible to recover the structure of a 500kD molecule to 0.3nm, a 1MD molecule to 0.4nm, and a 2MD molecule to 0.5nm [8]. We expect algorithmic improvements to extend the object size and achievable resolution further (see section 5.3 below). In summary, using noise-robust manifold mapping techniques, the structure of an object can be determined to high resolution from ultra-low-signal snapshots of identical members of an ensemble, each in an unknown orientation.

## 4. Capabilities of Manifold Mapping Techniques

### 4.1. *Dealing with Experimental Data*

Using simulated data, we have established that our approach is robust against the addition of the background measured in single-particle experiments at the FLASH FEL facility in Hamburg, Germany. As shown in figure 3, the R-factor of the reconstructed diffraction volume degrades with increasing background noise, but this can be reversed by increasing the number of snapshots provided to the algorithm. Using the Advanced Photon Source and a nanofoam as an analogue for a complex biological object, we have established that our approach can orient experimental snapshots to within a Shannon angle [8] down to the lowest scattered photon intensity measured to date (0.08 photon/pixel) [43]. Experiments at lower fluxes and similar studies on cryo-EM data are under way. These results further support the notion that manifold mapping algorithms can deal with experimental data.

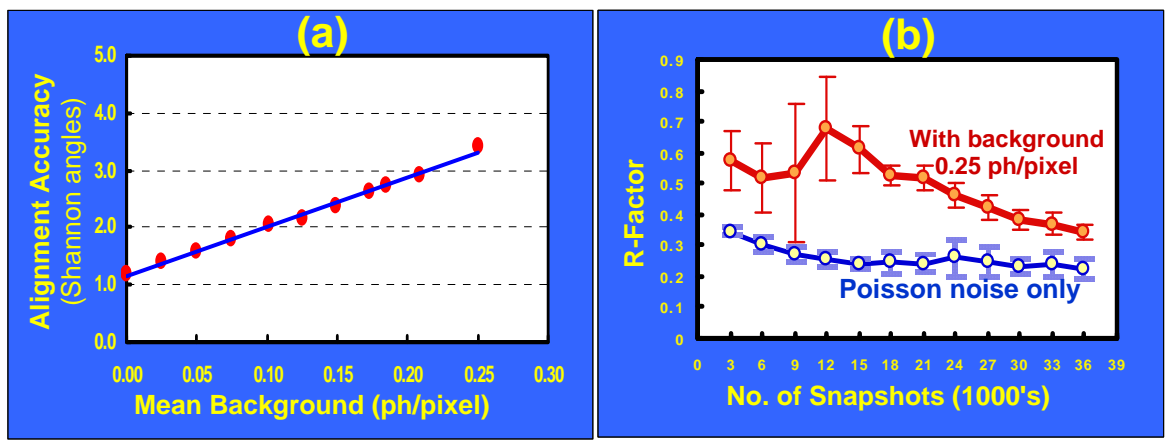

**Figure 3. Adding background scattering increases the R-factor, but this can be reversed by providing more snapshots to the algorithm.** (a) Alignment accuracy degrades with increasing background scattering. (b) R-factor between the correct diffraction volume and that reconstructed in the presence of background (red curve) tends toward that without background (blue curve) as the number of snapshots is increased. Signal: 0.04 photon/pixel@0.18nm with Poisson noise; background of 0.25 ph/pixel: ~1/3 of that measured at FLASH UV XFEL.

### 4.2. *Superior Signal Extraction*

Most manifold mapping approaches exploit the information content of the entire dataset to determine the manifold. GTM has the additional capability to generate a snapshot corresponding to any specified point on the manifold - hence the "generative" designation. The reconstructed image is not simply an average over the snapshots assigned to an orientational bin as in standard classification approaches [44,45], but stems from the entire data set. This produces signal extraction capabilities superior to approaches which classify images and then form class averages to reduce noise. To illustrate this and make contact with cryo-EM, we use real-space, rather than diffraction snapshots. GTM was used to sort into 60 orientational bins 3600 simulated electron microscope images of the small protein chignolin in random orientations (figure 4). The snapshots were simulated at a mean electron count of $10^2/\text{Å}^2$ at a point-to-point resolution of 0.37nm (Phillips CM20T TEM; 100kV; Scherzer defocus) (figure 4(b)). Like other orientation recovery techniques, GTM was able to accurately classify the snapshots into orientational bins. At this point, standard approaches average the snapshots within a bin to reduce noise. This produces the images shown in figure 4(c). The images generated by GTM from the manifold are shown in figure 4(d). Their improved quality is a visual demonstration of the superior signal extraction capability of manifold mapping techniques.

It is not straightforward to relate the mean electron dose used in figure 4 to those typically used in cryo-EM. Using the ratio {image variance/(mean electron dose)$^2$} as a measure of contrast, we compared the

- 6 -

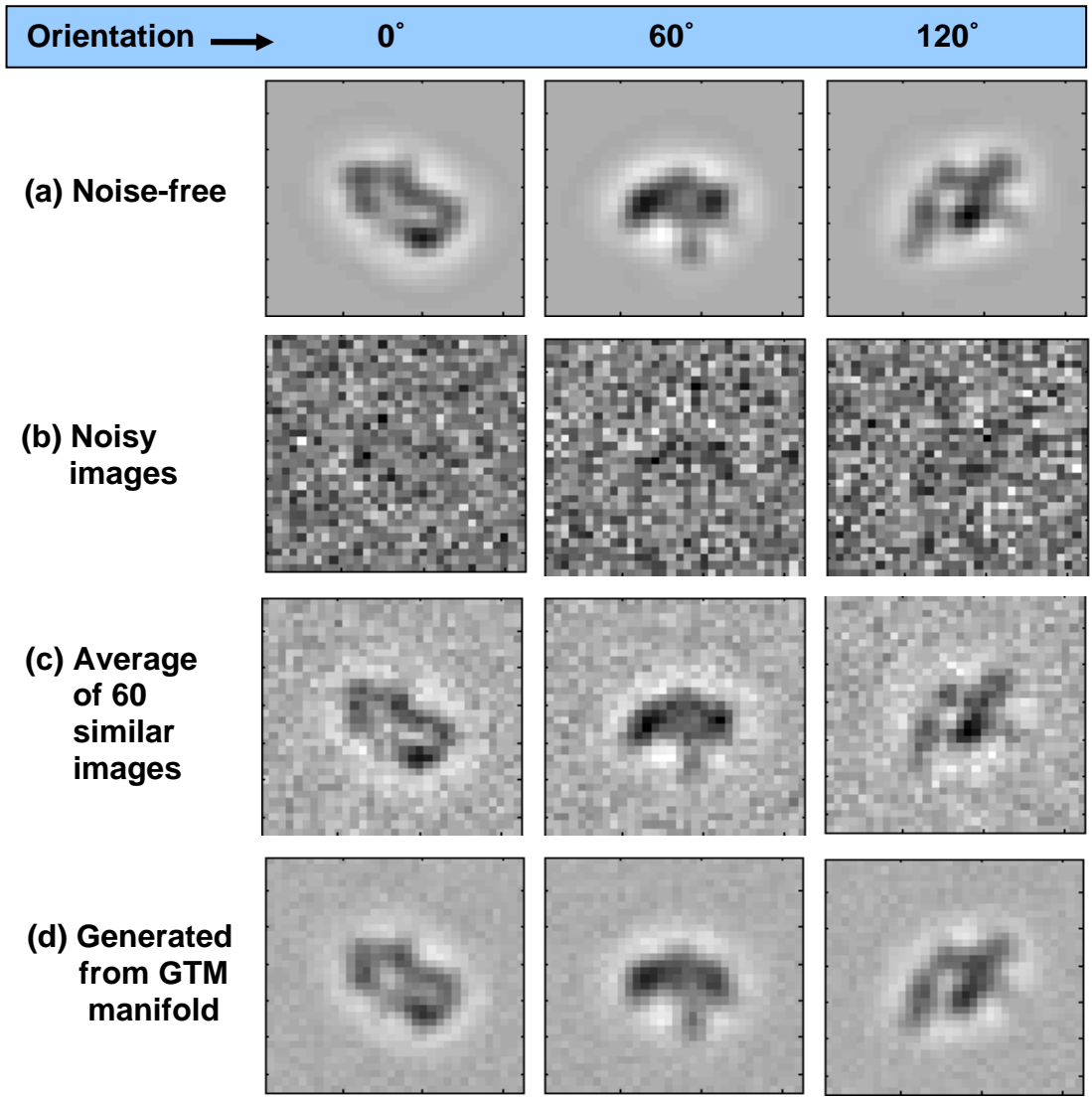

**Figure 4. Generating images from the manifold gives superior results.** Simulated cryo-EM images of the protein chignolin at 0.37nm resolution in three orientations. (a) Noise-free snapshots. (b) Snapshots at a mean electron count of $10^2$ electrons/Å$^2$. (c) Images obtained by averaging the snapshots assigned to each orientational bin. (d) Images generated from the GTM manifold. Note the superior signal-to-noise ratio when the image is generated from the manifold.

simulated images of chignolin at $10^2$ electrons/Å$^2$ (figure 4) with those of GroEL at 32 electrons/Å$^2$, the dose used in a recent cryo-EM study [6]. The differences in the contrasts and symmetries of the two particles indicate that the dose of figure 4 is roughly an order of magnitude lower than that commonly used in cryo-EM. This estimate requires validation with experimental data, because effects such as absorption contrast and substrate noise cannot be easily simulated. Nevertheless, the enhanced performance of manifold mapping over conventional image classification and averaging is enticing. Generation of manifold-based images with other manifold mapping techniques [36,38] is in principle possible, but has not been demonstrated. The ability to use the information content of the entire dataset for each reconstructed image is of decisive importance for recovering orientation, structure, and conformations, both in cryo-EM and XFEL.

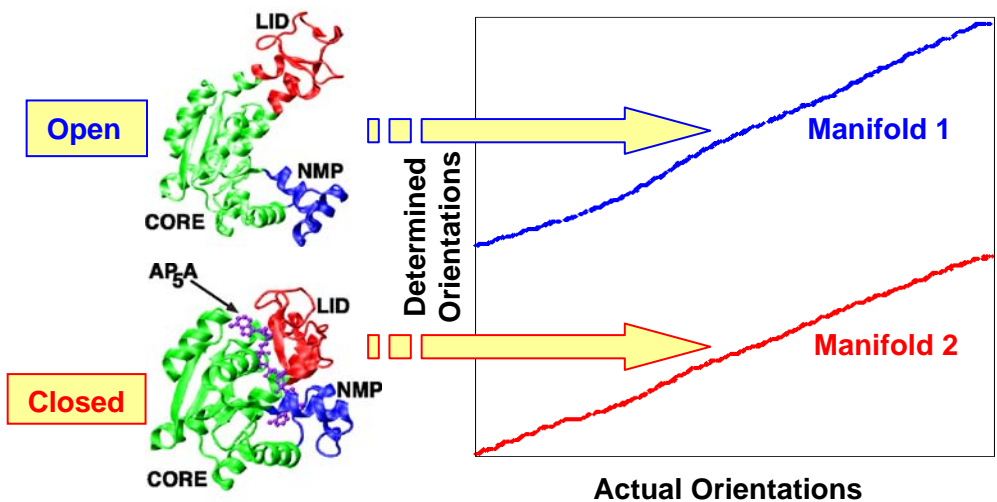

**Figure 5. Manifold mapping separates snapshots from different conformations and find the orientations within each set with no a priori knowledge.** When a mixture of diffraction snapshots from the molecule ADK in its open (Protein databank (PDB) entry 4AKE) and closed (PDB entry 1ANK) conformations is presented to noise-robust versions of GTM or Isomap at signal levels corresponding to 0.04 ph/pixel@0.18nm, the snapshots are automatically sorted into different manifolds and their orientations determined. The results shown here stem from Isomap analysis, with GTM results being very similar. The 8.5-$\sigma$ separation between the two manifolds implies extreme fidelity in separating different conformations.

## 5. Mapping Conformations

### 5.1. *Discrete Conformations and "Post facto Purification"*

We now show that when the bioparticle beam in an XFEL experiment contains different discrete conformations, manifold mapping automatically sorts the diffraction snapshots into separate classes and determines their orientations. This approach, also applicable to cryo-EM images, allows one to reconstruct the 3D structures of different conformations - and, by extension, species – separately. Figure 5 shows the results when a mixture of randomly oriented diffraction snapshots from the closed and open conformations of the molecule adenylate kinase (ADK; Protein Data Bank entries: 1ANK and 4AKE, respectively) are presented to noise-robust manifold mapping algorithms at a signal level corresponding to $4 \times 10^{-2}$ photons/pixel at 0.18nm. Because of their chemical identity, the conformations of ADK are extremely difficult to separate chemically. As shown in Figure 5, manifold mapping (by GTM or Isomap) automatically sorts the snapshots into separate manifolds, and determines their orientations to within a Shannon angle. We note that no prior information was provided to the algorithm regarding the types or number of conformations.

The confidence level with which sorting was performed can be deduced as follows. Noise causes the vectors representing the snapshots to depart from the noise-free manifolds, thus giving a certain "thickness" to each manifold. The sorting confidence can be deduced from the closest separation between the two manifolds expressed in standard deviations of the distributions of vectors about the manifolds. At the signal level of $4 \times 10^{-2}$ photon/pixel at 0.18nm with Poisson noise, the smallest separation between the two manifolds exceeds 8.5 standard deviations. This means that snapshots from the different conformations are sorted with extreme fidelity. We note that larger objects such as macromolecular assemblies produce larger signals. It should therefore be possible to use manifold techniques to map their conformations with even greater precision. Manifold mapping can sort discrete molecular conformations - and, by extension, different species - with extreme fidelity. This offers possibility to use solutions containing multiple species and perform "post facto purification" by sorting the data after the experiment.

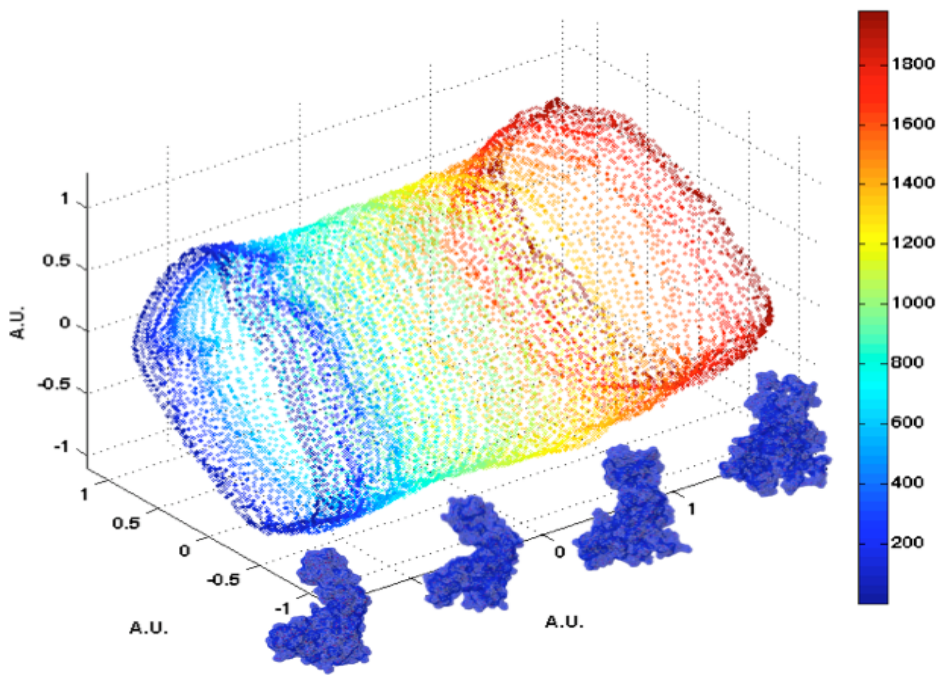

**Figure 6. Manifold traced out by 100 conformations of an unfolding ADK molecule, each able to assume one of 125 orientations about a single axis.** The molecular conformations corresponding to different points on the manifold are shown in the insets.

### 5.2. *Conformational Continua*

Ultimately, one would like to map the entire continuum of conformations assumed by a biological system. We use the unfolding of a molecule to demonstrate the principle of mapping conformational continua. The unfolding of ADK was simulated by molecular dynamics as follows. The coordinates of ADK from *E. coli* in the open state (Protein Data Bank entry: 4AKE) were placed in a spherical droplet of water and simulated at a nominal temperature of 850 K for 5 ns using NAMD [46]. 12,500 diffraction snapshots were simulated from 100 conformations, with each conformation assuming 125 orientations about one axis. Snapshots were provided to a modified version of the Isomap manifold mapping algorithm, and the resulting manifold displayed through its projections along the first three dominant eigenvectors obtained by Isomap analysis (figure 6). It is clear that orientational and conformational variations combine to produce a tubular manifold. Qualitatively, the closed cross-sections of the tube include orientational change, while paths terminating at the tube ends indicate conformational change. In order to separate an orientational change from a conformational change, however, the directions corresponding to pure orientational and pure conformational change must be identified at each point on the manifold. This can be achieved by recognizing that the manifold is Riemannian. Due to the SO(3) symmetry of molecular orientations, the Killing vectors on the manifold point in directions of pure orientational change, and thus also identify the directions of pure conformational change. Manifolds with SO(3) symmetry in some directions have received considerable attention in General Relativity and lattice space-time [47], and well-established techniques exist for determining their Killing vector fields. Noise-robust versions of Riemannian techniques can be used to identify directions of pure orientational and conformational change on the manifold, and thus reconstruct the 3D structure of the conformations of macromolecular assemblies. A demonstration of this approach is underway.

### 5.3. *Computational Issues*
The computational demands of current algorithms rise rapidly with the number of resolution elements $r$, typically as $r^n$ with $2 \leq n \leq 3$, where $r \sim (D/d)^3$, and $D$ and $d$ designate the object diameter and resolution, respectively [8,32]. The discretization of a conformational continuum into $c$ points increases the computational load by the same factor. We have taken the following steps to reduce the computational requirements of GTM. a) By processing data in small batches, the memory requirements have been reduced to 2GB of RAM, which is available at each core. b) A modified approach based on a binary search has been developed, which should reduce the exponent $n$ by 1, and the computational load for a typical calculation by $\sim 10^6$x. This is in implementation. c) By insuring that the modified code can be straightforwardly parallelized, we expect to reduce the computational time by another 100x. Isomap and Diffusion Map are computationally less expensive than GTM. We have identified significant load reduction measures for these techniques also. These algorithmic modifications help insure that the increased computational load due to conformational variability can be handled.

### 5.4. *Data Collection*
In the absence of conformational variability, structure recovery to high resolution can require $\sim 10^6$ snapshots. Ignoring the mutual information between different snapshots (which reduces the number of snapshots needed), mapping a conformational continuum discretized into 100 points requires 100x more data. The LCLS source and relevant detector operate at 120Hz and thus able to deliver $10^7$ shots per day. The planned European XFEL is slated to operate at 30kHz. It is thus realistic to expect that the data needed to map conformational continua can be indeed collected.

## 6. Discussion
The majority of algorithms used in cryo-EM and those originally proposed for XFEL single-particle work sort similar snapshots into orientational classes, average over members of each class to boost signal, and finally determine orientations. Three years ago, Shneerson et al. [9] showed that structure recovery by such algorithms requires 1000x more flux than theoretically possible with currently envisaged XFELs. Since then it has become clear that single-particle structure recovery, be it from XFEL diffraction snapshots or cryo-EM images, is most efficiently performed with approaches which exploit the information content of the entire dataset [5,8,13]. Analysis of simulated data indicates that such approaches can recover single-molecule structure at anticipated XFEL fluxes, and perhaps at doses lower than currently in use in cryo-EM.

These emerging algorithms can be broadly separated into three classes: (i) Information theoretic; (ii) Manifold Embedding; and (iii) Manifold Mapping. The information theoretic approaches such as GTM use Bayesian inference and an expectation-maximization engine with constraints to insure contiguity of the manifold in data space. This can be a specified latent space of orientations [8,35] or a less familiar equivalent, such as the "expansion-contraction" cycle used in [32]. In essence, these algorithms are "general" in the sense that they contain no information about the diffraction process. For this reason, they are computationally inefficient [48]. The manifold embedding approaches such as Diffusion Map and Isomap use graph theoretic means to discover the manifold in data space without recourse to a latent space. They also incorporate no information about the diffraction process. While, their kernel-based approach is computationally more efficient, they are noise-sensitive and unable to identify the correct manifold topology without prior denoising [39]. Finally, manifold mapping techniques circumvent embedding altogether, using Riemann geometric techniques to "navigate on the manifold directly" without recourse to external coordinate systems. The nature of the diffraction process can be directly incorporated so as to place constraints on the mapping and the properties of the data manifold [49]. As such, they promise an optimal combination of computational efficiency and noise robustness, although this remains to be demonstrated.

Graph-theoretic and Riemann geometric approaches are perhaps most appropriate for studying conformational continua, because they require no a priori knowledge of the dimensionality of the data manifold. In contrast, Bayesian expectation-maximization-based approaches either require prior knowledge of the number of conformations present, or must deduce this by trial and error [5]. More generally, Riemann geometric manifold mapping approaches offer a fundamentally new formulation of scattering with potential access to an array of powerful analytical techniques. The implications of this remain to be fully understood. It is nonetheless intriguing that methods developed in cosmology to study the structure of the universe may be used to investigate the building blocks of life.

## 7. Summary and Conclusions

We have outlined powerful new algorithmic approaches based on concepts from information theory, graph theory, and Riemannian geometry, and demonstrated their potential for extracting signal from noise, recovering 3D structure with no orientational information, separating discrete conformations and species, and eventually mapping conformational continua. By naturally incorporating conformational heterogeneity, these algorithms promise to substantially increase the range of systems accessible to single-particle techniques such as cryo-EM, and emerging X-ray Free-Electron Laser (XFEL)-based approaches. At the same time, these algorithms offer a route to transforming fundamental limits due to radiation damage and noise to computational issues, because, down to some as-yet-undetermined limit, signal can be traded against the number of snapshots. As computational resources have increased exponentially for five decades, such a development would constitute a fundamental advance in structure recovery.

**Acknowledgements**
We are grateful to Dimitris Giannakis, Leonard Parker, Dilano Saldin, and Valentin Shneerson, for valuable discussions, and to Ross Harder and Ian McNulty for providing experimental nanofoam data for our analysis. This work was supported in part by NIH Protein Structure initiative grant GM074901.